\documentclass[]{relaxed_system_lab}

\usepackage[utf8]{inputenc}
\usepackage[T1]{fontenc}
\usepackage{geometry}
\usepackage{amsmath,amssymb}  %
\usepackage{charter}

\usepackage[toc,page,header]{appendix}

\usepackage{algorithm}
\usepackage{array}
\usepackage{wrapfig}
\usepackage{amsmath}
\usepackage[utf8]{inputenc} % allow utf-8 input
\usepackage[T1]{fontenc}    % use 8-bit T1 fonts
\usepackage{hyperref}       % hyperlinks
\usepackage{url}            % simple URL typesetting
\usepackage{booktabs}       % professional-quality tables
\usepackage{amsfonts}       % blackboard math symbols
\usepackage[most]{tcolorbox}
\usepackage{nicefrac}       % compact symbols for 1/2, etc.
\usepackage{microtype}      % microtypography
\usepackage{xcolor} 
\usepackage{amssymb}
\usepackage{caption}
\usepackage{xspace}
\usepackage{multirow}
\usepackage{xcolor}
\usepackage{graphicx} 

\usepackage{xspace}
\usepackage{multirow}
\usepackage{seqsplit}

\usepackage{tcolorbox}
\usepackage{mdframed}
\mdfdefinestyle{insightbox}{
  backgroundcolor=green!5!white,
  linecolor=green!40!black,
  linewidth=0.3pt,
  roundcorner=2pt,
  innerleftmargin=4pt,
  innerrightmargin=4pt,
  innertopmargin=3pt,
  innerbottommargin=3pt
}
\mdfdefinestyle{motivationbox}{
  backgroundcolor=blue!5!white,
  linecolor=blue!40!black,
  linewidth=0.3pt,
  roundcorner=2pt,
  innerleftmargin=4pt,
  innerrightmargin=4pt,
  innertopmargin=3pt,
  innerbottommargin=3pt
}
\mdfdefinestyle{codebox}{
  backgroundcolor=black!4!white,
  linecolor=black!40!black,
  linewidth=0.3pt,
  roundcorner=2pt,
  innerleftmargin=4pt,
  innerrightmargin=4pt,
  innertopmargin=3pt,
  innerbottommargin=3pt
}

\usepackage[utf8]{inputenc} % allow utf-8 input
\usepackage[T1]{fontenc}    % use 8-bit T1 fonts
\PassOptionsToPackage{hyphens}{url}
\usepackage{hyperref}       % hyperlinks
\usepackage{xurl}           % URL typesetting with flexible line breaks
\usepackage{booktabs}       % professional-quality tables
\usepackage{amsfonts}       % blackboard math symbols
\usepackage{nicefrac}       % compact symbols for 1/2, etc.
\usepackage{microtype}      % microtypography
\usepackage{xcolor}         % 
\usepackage{xspace}
\usepackage{amssymb}
\usepackage{tabularx}
\usepackage{caption}

\newcommand{\name}{\textsc{HexiSeq}\xspace}

\usepackage{amsmath}
\usepackage{algorithm}
\usepackage{algpseudocode}
\usepackage{graphicx}
\usepackage{subcaption}
\usepackage{wrapfig}

\ifdefined\final
\usepackage[disable]{todonotes}
\else
\usepackage[textsize=tiny]{todonotes}
\fi

\usepackage{natbib}
\usepackage{latexsym}

\usepackage{url}
\usepackage{amssymb}
\usepackage[utf8]{inputenc}
\usepackage{microtype}
\usepackage{booktabs}
\usepackage{pifont} 
\usepackage{multirow}
\usepackage{makecell}
\usepackage{paralist}
\usepackage{xspace}
\usepackage{color}
\usepackage{xcolor}
\usepackage{colortbl}
\usepackage{adjustbox}
\usepackage{hyperref} 
\usepackage[edges]{forest}
\usepackage{tikz} 
\usepackage{caption}
\usepackage{amsfonts}

\hypersetup{
    colorlinks,
    linkcolor={blue!80!black},
    citecolor={blue!80!black},
}
\tikzset{
    root/.style =             {align=center, text width=1cm, rounded corners=3pt, line width=0.3mm, fill=gray!10, draw=gray!80, font=\small},
    demographic/.style =         {align=center, text width=1.8cm, rounded corners=3pt, line width=0.3mm, fill=blue!10, draw=blue!80, font=\footnotesize},
    demographic_work/.style =    {align=center, text width=10cm, rounded corners=3pt, line width=0.3mm, fill=blue!10, draw=blue!0, font=\footnotesize},
    character/.style =         {align=center, text width=1.8cm, rounded corners=3pt, line width=0.3mm, fill=red!10, draw=red!80, font=\footnotesize},
    character_work/.style =    {align=center, text width=10cm, rounded corners=3pt, line width=0.3mm, fill=red!10, draw=red!0, font=\footnotesize},
    personalization/.style =           {align=center, text width=1.8cm, rounded corners=3pt, line width=0.3mm, fill=cyan!10, draw=cyan!80, font=\footnotesize},
    personalization_work/.style =      {align=center, text width=10cm, rounded corners=3pt, line width=0.3mm, fill=cyan!10, draw=cyan!0, font=\footnotesize},
    risk/.style =         {align=center, text width=1.8cm, rounded corners=3pt, line width=0.3mm, fill=orange!10, draw=orange!80, font=\footnotesize},
    risk_work/.style =    {align=center, text width=10cm, rounded corners=3pt, line width=0.3mm, fill=orange!10, draw=orange!0, font=\footnotesize},
}

\usepackage{CJK}

\newtcolorbox{promptbox}[1][]{
  enhanced,
  breakable,
  colback=promptboxlightgray,
  colframe=promptboxblue!30,
  arc=8pt,
  boxrule=0.5pt,
  left=12pt,
  right=12pt,
  top=8pt,
  bottom=8pt,
  fonttitle=\bfseries,
  fontupper=\linespread{1.2}\selectfont,
  title=#1
}

\title{\name: Accommodating Long Context Training of LLMs over Heterogeneous Hardware}

\author{Yan Liang$^1$, Youhe Jiang$^1$, Ran Yan$^1$, Binhang Yuan$^1$, Wei Wang$^1$, Chuan Wu$^2$}

\affiliation{$^1$The Hong Kong University of Science and Technology, $^2$The University of Hong Kong}

\abstract{
Long-context training of large language models (LLMs) is commonly distributed with Context Parallelism (CP) and Head Parallelism (HP), but existing training systems largely assume homogeneous GPU meshes. This paper extends CP and HP to heterogeneous GPU clusters with mixed GPU models and non-uniform network bandwidths, a common setting in production training. We introduce \name, a system that supports fully asymmetric CP--HP partitioning by assigning sequence shards and attention heads according to device compute, memory, and communication capabilities. We formalize heterogeneous CP--HP allocation as a constrained optimization problem and develop an efficient hierarchical scheduler for finding optimal schedules. We evaluate \name against state-of-the-art CP and HP baselines on both real and simulated heterogeneous clusters. Across models from 3B to 70B parameters and context lengths up to one million tokens, \name improves throughput by $1.11\times$ on average and up to $1.19\times$ on mixed H100--A100 testbeds, and by $1.36\times$ on average and up to $1.72\times$ in simulations with 32--128 GPUs spanning up to four GPU models. On FLOP-comparable pairs against homogeneous clusters, \name reaches throughput close to the strongest homogeneous baseline, showing that heterogeneous clusters can be used efficiently for long-context LLM training.
}

\begin{document}

\maketitle

\section{Introduction}
\label{sec:introduction}
Long-context training has become a foundational workload for large language models (LLMs), enabling applications such as long-document analysis, repository-scale code reasoning, and extended multi-turn interaction. Frontier models such as GPT-5.5~\citep{openai2026gpt55}, Claude Opus 4.7~\citep{anthropic2026claudeopus47}, and Gemini 3.1 Pro~\citep{google2026gemini31pro} now expose context windows at the million-token scale. Supporting such capabilities requires training systems to process extremely long sequences, where attention computation grows quadratically with sequence length and activation memory grows rapidly with context size. However, existing long-context parallel training systems typically assume homogeneous GPU meshes, with identical devices, uniform sequence shards, uniform head placement, and symmetric communication groups. As a result, scaling million-token training often depends on provisioning homogeneous clusters of high-calibre GPUs, whose cost and availability limit broader long-context LLM development. This paper explores an alternative approach: distributing long-context training across heterogeneous GPU clusters, thereby improving resource utilization and expanding access to large-context training.

Heterogeneous clusters are increasingly common in practice. As AI accelerators evolve rapidly, new hardware generations are introduced every few years, while older generations often remain in service for much longer. For example, NVIDIA has successively introduced A100, H100, B200, and B300 accelerators from 2020 to 2025~\citep{nvidia2020ampere,nvidia2022hopper,nvidia2024blackwell,nvidia2025dgxb300}. In practice, these newer devices are often deployed alongside earlier accelerators such as V100, A800, and L40S in clouds, industrial labs, and academic research clusters. This coexistence creates a widely available yet heterogeneous hardware pool and highlights an opportunity to use existing mixed-generation resources for cost-effective long-context LLM training. As context length grows, the attention workload becomes both larger and more divisible, creating more room for heterogeneous CP and HP schedules to utilize compute, memory, and communication capacities across mixed accelerator models.

Realizing this opportunity, however, is challenging from both scheduling and system implementation perspectives. Long-context training systems typically combine IO-aware attention kernels such as FlashAttention~\citep{dao2022flashattention,dao2024flashattention} with Context Parallelism (CP) and Head Parallelism (HP). CP partitions the sequence dimension and circulates key-value (KV) blocks through Ring Attention~\citep{liuringattention}, while HP redistributes attention heads through all-to-all (A2A) collectives in the style of Ulysses~\citep{jacobs2024system}. Hybrid systems such as USP and LoongTrain~\citep{fang2024usp,gu2024loongtrain} compose CP and HP into a single training mesh. These systems, however, largely assume homogeneous GPU configurations: A2A groups have the same cardinality, GPU ranks within a group hold equal-length sequence shards, and attention heads are assigned uniformly across ranks. Such symmetry implicitly assumes that all GPUs perform attention computation at the same rate and exchange KV blocks over links with comparable bandwidth. This assumption breaks down on heterogeneous clusters, where devices differ in peak FLOPS, HBM (High Bandwidth Memory) capacity, HBM bandwidth, and pairwise communication bandwidth across both intra-node and inter-node links.

\nocite{peng2025hexgen,jiang2023hexgen,jiang2025hexgen,jiang2025demystifying,jiang2025thunderserve,tong2025parallax,jiang2026boute,jiang2026oserve,jiang2025cascadia,zhang2025efficient}

Concretely, heterogeneous long-context CP and HP training faces two fundamental challenges:

\textbf{\underline{C1:} Mismatched compute and memory capacities.} GPUs from different generations vary in peak FLOPS, HBM capacity, and HBM bandwidth. A symmetric CP and HP mesh assigns every rank the same sequence length and the same number of attention heads. Consequently, faster devices stall behind slower ones, while memory-rich devices cannot absorb the longer sequence shards their HBM capacity would otherwise support. Efficient heterogeneous training therefore requires partitioning sequence and head workloads according to each device's compute and memory profile.

\textbf{\underline{C2:} Non-uniform communication bandwidth.} Inter-device links differ widely, ranging from high-bandwidth NVLink within a node to lower-bandwidth RDMA, InfiniBand, or Ethernet across nodes. Long-context attention is highly sensitive to these differences because both A2A redistribution and Ring Attention KV exchange lie on the critical path. The slowest link in an A2A group can dominate redistribution latency, and the slowest hop on a ring can stall each Ring Attention step. Efficient execution therefore requires aligning A2A groups and KV exchange topology with the underlying communication fabric.

Existing systems fail to address these challenges. CP and HP frameworks such as USP and LoongTrain~\citep{fang2024usp,gu2024loongtrain} adopt symmetric partitioning strategies that under-utilize heterogeneous hardware. Conversely, heterogeneity-aware training systems such as HexiScale~\citep{yan2024hexiscale} and Cephalo~\citep{guo2025cephalo} support heterogeneous data, pipeline, and tensor model parallelism, but leave the CP and HP dimensions, which are central to long-context training, largely unaddressed. To bridge this gap, we propose \name, a system that exposes heterogeneity directly to CP and HP scheduling. Instead of fitting heterogeneous devices into a symmetric mesh, \name jointly schedules A2A grouping, sequence placement, and head placement according to device compute, memory, and communication capabilities.

Our contributions are summarized as follows.

\textbf{\underline{Contribution 1.}}
We implement \name, a heterogeneous long-context LLM training system that supports asymmetric CP and HP execution. \name introduces a schedule abstraction with variable A2A groups, non-uniform group-level and rank-level sequence shards, and weighted attention-head placement. Its runtime materializes heterogeneous process groups, performs ragged A2A redistribution with non-uniform split sizes, and decomposes Ring Attention into sub-ring KV exchange tasks while preserving the semantics of existing CP and HP attention.

\textbf{\underline{Contribution 2.}}
We formulate heterogeneous CP and HP scheduling as a constrained optimization problem over A2A grouping, sequence placement, and head placement, subject to per-device memory and inter-device bandwidth constraints. We develop an analytical performance model for non-attention computation, A2A redistribution, Ring Attention compute and communication, and activation memory. We use this model inside an efficient hierarchical scheduler that generates topology-aware A2A partitions, plans group-level sequence splits, and refines rank-level sequence shards and head counts.

\textbf{\underline{Contribution 3.}}
We evaluate \name on physical heterogeneous testbeds and larger simulated clusters. On mixed H100 and A100 testbeds with up to $16$ GPUs, covering $3$B, $7$B, and $13$B models and context lengths from $8$K to $256$K, \name achieves $1.11\times$ throughput on average and up to $1.19\times$ over the strongest of USP, Ulysses, and Ring Attention. In simulations with $32$ to $128$ GPUs spanning up to four accelerator generations, and with $13$B and $70$B models at context lengths up to $1024$K, \name improves throughput by $1.36\times$ on average and up to $1.72\times$ over the strongest heterogeneous baseline. Against homogeneous clusters with matched aggregate FLOPs, \name reaches comparable throughput while running on heterogeneous hardware.

\section{Related Work}
\label{sec:related_work}
\textbf{Long-context attention parallelism.}
Data parallelism (DP), tensor parallelism (TP), and pipeline parallelism (PP) distribute model parameters and layers in LLM training~\citep{li2020pytorch,rajbhandari2020zero,shoeybi2019megatron,narayanan2021efficient,huang2019gpipe,narayanan2019pipedream,li2021terapipe}, but they do not by themselves resolve the memory and communication pressure of a single long sequence. Context-parallel systems partition the sequence dimension and circulate KV blocks through Ring Attention~\citep{liuringattention,li2023sequence}, while head-parallel systems such as Ulysses use A2A collectives to redistribute attention heads~\citep{jacobs2024system}. Hybrid designs, including USP and LoongTrain, combine both axes to scale to longer contexts~\citep{fang2024usp,gu2024loongtrain}. A second line of work improves routing, overlap, and memory scheduling for the same attention-parallel patterns~\citep{korthikanti2023reducing,brandon2023striped,lidistflashattn,chen2024internevo,sun2024burstattention,luo2024mini,liustartrail,chen2025mesh,chen2025sppo,zhao2025dsp,zhao2025memo,yao2025training}. These systems largely assume that all ranks within the attention mesh are symmetric and exchangeable. \name preserves their attention semantics, but removes this uniformity assumption by allowing sequence shards, head counts, and A2A groups to match heterogeneous hardware.

\textbf{Heterogeneous distributed training.}
A growing line of systems balances training across devices with different compute or memory capacities. HexiScale~\citep{yan2024hexiscale} adapts data, pipeline, and tensor model parallel placement to mixed GPUs; Poplar~\citep{zhang2025poplar} pursues similar adaptation; and broader systems extend heterogeneity-aware scheduling to multi-region and geo-distributed settings~\citep{strati2025sailor,tang2025h2,wang2025deepcee,li2025hetu,guo2025cephalo,obeidi2026heterogeneous}. These works show that asymmetric scheduling is essential when hardware differs, but they schedule at the granularity of samples, layers, tensor shards, or pipeline stages, and none of them parameterise the context- or head-parallel dimensions. As a consequence, in the long-context regime that this paper targets, a heterogeneous DP/PP/TP plan cannot avoid out-of-memory failures on the smaller-HBM partition without falling back to a CP layer. \name targets this attention-internal structure rather than full-sample, layer, or pipeline-stage placement, making it complementary to the systems above.

\nocite{miao2022galvatron,jiang2022osdp,wang2024improving,he2026efficient,yan2025areal,yan2025fsa}

\textbf{Variable-length and imbalance-aware training.}
A complementary line of work addresses workload imbalance caused by non-uniform sequence lengths in the training data~\citep{ge2024enabling,jiang2025dcp,ge2025bytescale,wang2025flexsp,li2025hydraulis}. These methods rebalance data-level work by changing how examples, tokens, or variable-length sequences are packed and distributed across ranks. \name instead addresses hardware-level imbalance for a fixed long-context training workload: the question is not how to batch uneven examples, but how to partition one long sequence and its attention heads across devices with different compute, memory, and communication capabilities.

\section{System Design and Implementation}
\label{sec:system_design}
This section presents \name's system design and implementation. We first use a representative heterogeneous cluster to compare several parallelization strategies and identify why symmetric CP and HP meshes are inefficient on such hardware (\S\ref{subsec:case_study}). We then describe the asymmetric parallel support that \name introduces to remove these limitations (\S\ref{subsec:asymmetric_support}).

\subsection{Case Study: Parallelism over Heterogeneity}
\label{subsec:case_study}

Consider training a 7B LLM with long-context inputs on a heterogeneous 8-GPU cluster consisting of one 2-GPU H100 node, one 2-GPU A100 node, and one 4-GPU A800 node. GPUs within each node are connected by high-bandwidth NVLink, while GPUs across nodes communicate through lower-bandwidth RDMA. Below, we compare three symmetric parallel strategies on this cluster with \name's asymmetric execution.
    
\textbf{Training with a symmetric CP and HP mesh.}
Existing CP and HP systems enforce a symmetric layout: each A2A group has identical cardinality, each rank holds an equal-sized sequence shard, and attention heads are uniformly distributed. Under this constraint, three representative execution plans arise, each incurring a distinct communication bottleneck:

\begin{figure}{t}
\centering
\vspace{-10pt}
\includegraphics[width=0.8\linewidth]{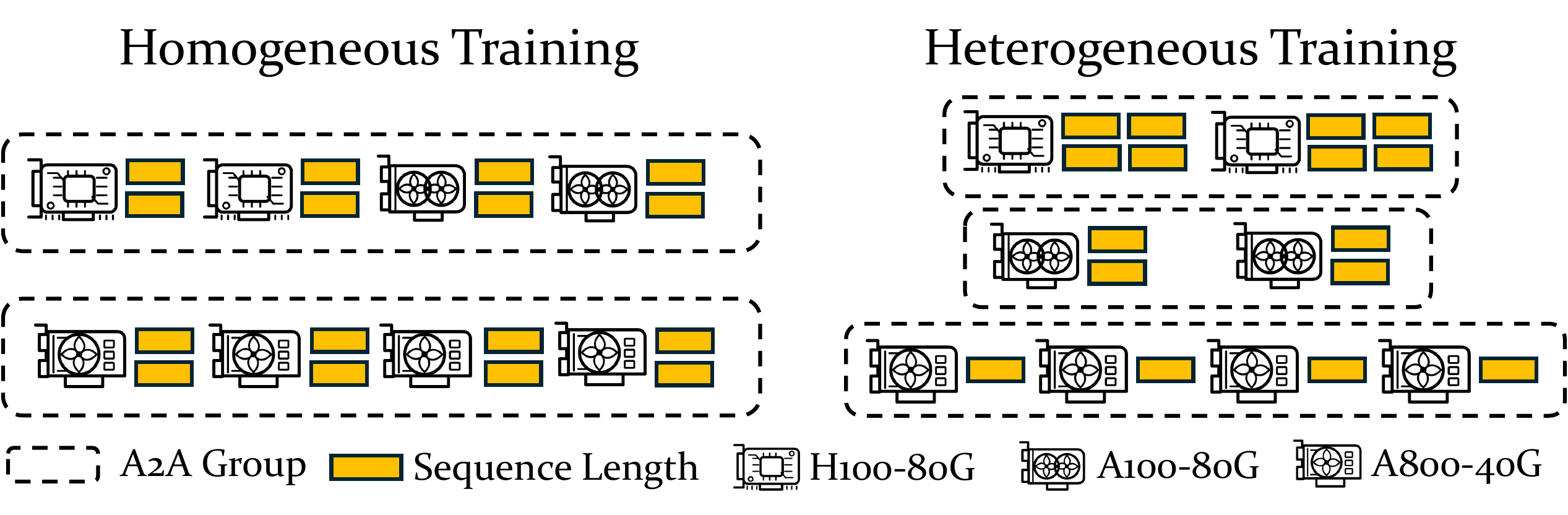}
\caption{Motivating load imbalance in long-context attention on heterogeneous clusters.}
\label{fig:motivation}
\end{figure}

\begin{itemize}
    % \vspace{-0.5em}
    \item {\textbf{Plan~1: Ring Attention~\citep{liuringattention}.}
    The system organizes all ranks into a single $8$-rank CP ring that spans every node, and attention is computed through sequential KV exchanges along this ring.
    Since the ring traverses inter-node links, the latency of each pipeline stage is dominated by the slowest cross-node hop, which leaves the high-bandwidth intra-node links underutilized.}
    % \vspace{-0.25em}
    
    \item {\textbf{Plan~2: Ulysses-style A2A~\citep{li2023sequence}.}
    The system performs a global $8$-way A2A communication to redistribute attention heads across all ranks.
    Since every A2A phase entails full cross-node data exchange, overall throughput is bounded by inter-node bandwidth. Furthermore, the uniform allocation of heads and tokens forces heterogeneous devices (e.g., GPUs of different generations) to operate under identical loads, thereby precluding capability-proportional utilization.}
    % \vspace{-0.25em}
    
    \item {\textbf{Plan~3: USP-style with uniform allocation~\citep{fang2024usp}.}
    The system arranges ranks into a regular $m{\times}n$ CP--HP mesh (e.g., $4{\times}2$ or $2{\times}4$), coupling a CP ring with intra-group A2A communication.
    However, this structure exhibits complementary bottlenecks:
    (\textbf{\underline{i}}) in the $4{\times}2$ layout, each A2A group spans multiple nodes, so every redistribution step is constrained by inter-node bandwidth;
    (\textbf{\underline{ii}}) in the $2{\times}4$ layout, A2A communication remains intra-node, but the CP ring crosses node boundaries, causing each attention step to stall on the slowest inter-node link.}
    % \vspace{-0.5em}
\end{itemize}
    
    \textbf{Training with \name.}
    Unlike the uniform partitioning and symmetric communication structure of Plans~1--3, \name jointly accounts for network topology and device heterogeneity. Specifically, it schedules the same workload as three asymmetric A2A groups aligned with node boundaries as visualized in Figure~\ref{fig:motivation}. Within and across groups, sequence shards and head counts are weighted by device capability: H100s receive more work per device, A100s receive an intermediate load, and A800s receive smaller per-rank shards that fit their compute and memory profile.
    
    This schedule keeps A2A traffic inside high-bandwidth domains and assigns attention workload according to device capability, preserving CP and HP semantics while avoiding OOM on smaller-HBM devices and reducing stalls on faster devices. Producing such an asymmetric schedule requires both system support for non-uniform A2A groups, sequence shards, and head counts (\S\ref{subsec:asymmetric_support}), and a scheduler that can navigate the larger schedule space (\S\ref{sec:scheduler_design}).
    
    \subsection{Asymmetric Parallel Support in \name}
    \label{subsec:asymmetric_support}
    
    To train long-context LLMs efficiently on heterogeneous GPU clusters, we implement \name with fully asymmetric context-parallel and head-parallel execution. The key design is to separate the logical schedule from the attention runtime. Concretely, the scheduler emits a schedule $\mathcal{T}$ that specifies A2A groups, group-level sequence lengths, rank-level sequence shards, and per-rank head counts. The runtime materializes the corresponding process groups and communication primitives. The essential changes are as follows.

    \textbf{Schedule abstraction.}
    The schedule abstraction, formalized in \S\ref{subsec:formalization}, relaxes symmetric CP and HP execution along three axes: 
    (\textbf{\underline{i}}) allowing A2A groups to have different cardinalities, keeping bandwidth-heavy redistribution within high-bandwidth domains when beneficial; 
    (\textbf{\underline{ii}}) assigning the global sequence length unevenly across groups so that faster groups receive longer subsequences and reduce synchronization stalls; and 
    (\textbf{\underline{iii}}) refining intra-group placement by assigning per-rank sequence shards and head counts according to device capability while preserving attention semantics.

    \textbf{Heterogeneous A2A with per-rank split tables.}
    Conventional A2A-based CP implementations use symmetric redistribution, where sequence and head partitions are evenly split across ranks. This uniform placement under-utilizes faster devices in heterogeneous clusters, as all ranks receive comparable workloads regardless of their capabilities. \name replaces symmetric redistribution with heterogeneous ragged A2A, which directly supports non-uniform sequence--head mappings. Given a schedule, the runtime derives per-rank input and output split tables so that stronger devices receive larger shards and weaker devices receive smaller shards during A2A communication.
    
    \textbf{Sub-ring KV exchange with batched async transfers.}
    A symmetric Ring Attention pipeline assumes each rank holds the same head count, so one ring advances all heads in lockstep. Under \name's schedule, ranks may hold different and non-contiguous head ranges, so the runtime decomposes KV exchange into sub-ring tasks over contiguous head ranges. Batched asynchronous point-to-point transfers launch these tasks together at each step and overlap communication with local attention compute; non-contiguous head slices are packed into contiguous buffers before transfer.

    Together, these mechanisms make asymmetric CP and HP execution practical on heterogeneous hardware, enabling heterogeneity-aware placement without enforcing symmetric layouts. Combined with the hierarchical scheduler in \S\ref{sec:scheduler_design}, these system primitives form the complete \name framework, whose efficiency and effectiveness on heterogeneous clusters are evaluated in \S\ref{sec:experiments}.

\section{Scheduling with Heterogeneity}
\label{sec:scheduler_design}
This section formalizes heterogeneous CP and HP scheduling as a constrained optimization problem (\S\ref{subsec:formalization}), describes the analytical performance model that scores candidate schedules (\S\ref{subsec:cost_model}), and presents the hierarchical scheduler that navigates the schedule space (\S\ref{subsec:scheduler}).

\subsection{Formalization}
\label{subsec:formalization}

We work over a heterogeneous device set $\mathcal{D}$ with profiles $\{c_d, b_d, m_d\}_{d\in\mathcal{D}}$, where $c_d$ is compute throughput, $b_d$ is device memory bandwidth, and $m_d$ is HBM capacity, of device $d$. The communication topology is represented by a graph $\mathcal{G}=(\mathcal{D},\mathcal{E})$, whose edge bandwidth $B_{i,j}$ approximates effective pairwise communication bandwidth. Given a global sequence length $L_{\mathrm{tot}}$, a schedule $\mathcal{T}\in\mathfrak{T}$ specifies an A2A partition $\mathcal{P}=\{G_1,\ldots,G_K\}$, per-group sequence lengths $L_k$ with $\sum_k L_k=L_{\mathrm{tot}}$, per-rank pre-A2A sequence shards $s_d^{\mathrm{pre}}$ that sum to $L_k$ within each group, and per-rank head counts $n_d$ that collectively cover all $N_{\mathrm{heads}}$ attention heads (Table~\ref{tab:notation}).

We seek the schedule that minimizes per-iteration latency subject to per-device memory feasibility:
\begin{equation}
\mathcal{T}^{\star} \;=\; \arg\min_{\mathcal{T}\in\mathfrak{T}}\; T_{\mathrm{iter}}(\mathcal{T})
\quad \text{s.t.} \quad
\mathrm{Mem}(d;\mathcal{T}) \le m_d,\;\; \forall d\in\mathcal{D},
\label{eq:problem}
\end{equation}
where $T_{\mathrm{iter}}(\mathcal{T})$ is the iteration latency under $\mathcal{T}$ and $\mathrm{Mem}(d;\mathcal{T})$ is the peak HBM occupied on device $d$. Eq.~\eqref{eq:problem} is hard because the schedule space $\mathfrak{T}$ grows combinatorially with the number of devices, while latency is non-separable across topology, sequence placement, and head placement: the bottleneck A2A group, the slowest Ring Attention hop, and the most loaded device are coupled through the attention pipeline. \name addresses this with an analytical performance model (\S\ref{subsec:cost_model}) and a hierarchical scheduler (\S\ref{subsec:scheduler}) that exploits the problem structure.

\begin{table}
\centering
\caption{Core notation used in scheduler formalization.}
\label{tab:notation}
\footnotesize
\setlength{\tabcolsep}{2.5pt}
\renewcommand{\arraystretch}{0.9}
\begin{tabularx}{\linewidth}{@{}l | X | l | X@{}}
\hline
\textbf{Symbol} & \textbf{Description} & \textbf{Symbol} & \textbf{Description} \\
\hline
$d$ & GPU device index & $\mathcal{D}$ & Set of GPU devices \\
\hline
$\mathcal{G}$ & Communication topology graph & $\mathcal{E}$ & Edges in $\mathcal{G}$ \\
\hline
$\mathcal{T}$ & Candidate schedule & $\mathcal{T}^{\star}$ & Optimal schedule \\
\hline
$\mathfrak{T}$ & Schedule space & $\mathcal{P}$ & Partition into A2A groups \\
\hline
$G_k$ & A2A group & $K$ & Number of groups / Ring steps \\
\hline
$c_d$ & Compute throughput & $m_d$ & Memory capacity \\
\hline
$b_d$ & Memory bandwidth & $B_{i,j}$ & Pairwise communication bandwidth \\
\hline
$L_{\mathrm{tot}}$ & Global sequence length & $L_k$ & Sequence length assigned to $G_k$ \\
\hline
$s_d^{\mathrm{pre}}$ & Pre-A2A sequence shard & $n_d$ & Heads assigned to device $d$ \\
\hline
$N_{\mathrm{heads}}$ & Total attention heads & $\mathrm{Mem}(d;\mathcal{T})$ & Peak HBM usage \\
\hline
$T$ & Latency quantity & $M_1, M_2, I_{\max}$ & Pruning budgets and coord.\ desc.\ iteration cap \\\hline
\end{tabularx}
\vskip -0.05in
\end{table}

\subsection{Performance Model}
\label{subsec:cost_model}

Comparing schedules whose communication topology, sequence placement, and head placement all differ requires a fast and uniform cost estimate. \name therefore uses an analytical performance model that predicts iteration latency from device compute throughput, memory bandwidth, memory capacity, and pairwise communication bandwidth. The model does not simulate every kernel; it captures the dominant terms that determine the critical path of long-context CP and HP training. The schedule variables $L_k$, $s_d^{\mathrm{pre}}$, and $n_d$ jointly control non-attention compute, sequence--head redistribution, Ring Attention payloads, and peak activation memory.

\textbf{Critical-path latency.}
For each Transformer block, execution decomposes into three components:
(\textbf{\underline{i}}) non-attention computation, including projections, MLPs, and normalization;
(\textbf{\underline{ii}}) sequence--head redistribution through A2A communication; and
(\textbf{\underline{iii}}) Ring Attention, in which blockwise attention computation overlaps with KV exchange.
Since A2A groups execute concurrently, their latency is modeled by the slowest group. The block latency is estimated as the critical path:
\begin{equation}
T_{\mathrm{blk}}(\mathcal{T})
=
\max_{d \in \mathcal{D}} T_{\mathrm{nonattn}}(d;\mathcal{T})
+
\max_{k} T_{\mathrm{A2A}}(G_k;\mathcal{T})
+
\sum_{t=0}^{K-1} T_{\mathrm{step}}(t;\mathcal{T}),
\label{eq:total_time}
\end{equation}
Here, $T_{\mathrm{nonattn}}(d;\mathcal{T})$, $T_{\mathrm{A2A}}(G_k;\mathcal{T})$, and $T_{\mathrm{step}}(t;\mathcal{T})$ denote the non-attention latency, intra-group A2A latency, and Ring Attention step latency, respectively. The full iteration latency $T_{\mathrm{iter}}(\mathcal{T})$ is obtained by scaling $T_{\mathrm{blk}}(\mathcal{T})$ by the number of layers and microbatches, with forward/backward costs absorbed into the component coefficients; detailed equations are deferred to Appendix~\ref{app:cost_derivation}.

\textbf{Memory feasibility.}
The same model also serves as a hard feasibility filter: for each device it estimates peak activation and attention-buffer memory from $s_d^{\mathrm{pre}}$, $L_k$ for $d\in G_k$, and $n_d$, and any schedule that exceeds device capacity is rejected. This prevents the scheduler from selecting high-throughput plans that survive only by overflowing a memory-limited GPU.

\subsection{Hierarchical Scheduler}
\label{subsec:scheduler}

\begin{wrapfigure}[20]{r}{0.46\linewidth}
\centering
\vspace{-12pt}
\includegraphics[width=0.98\linewidth]{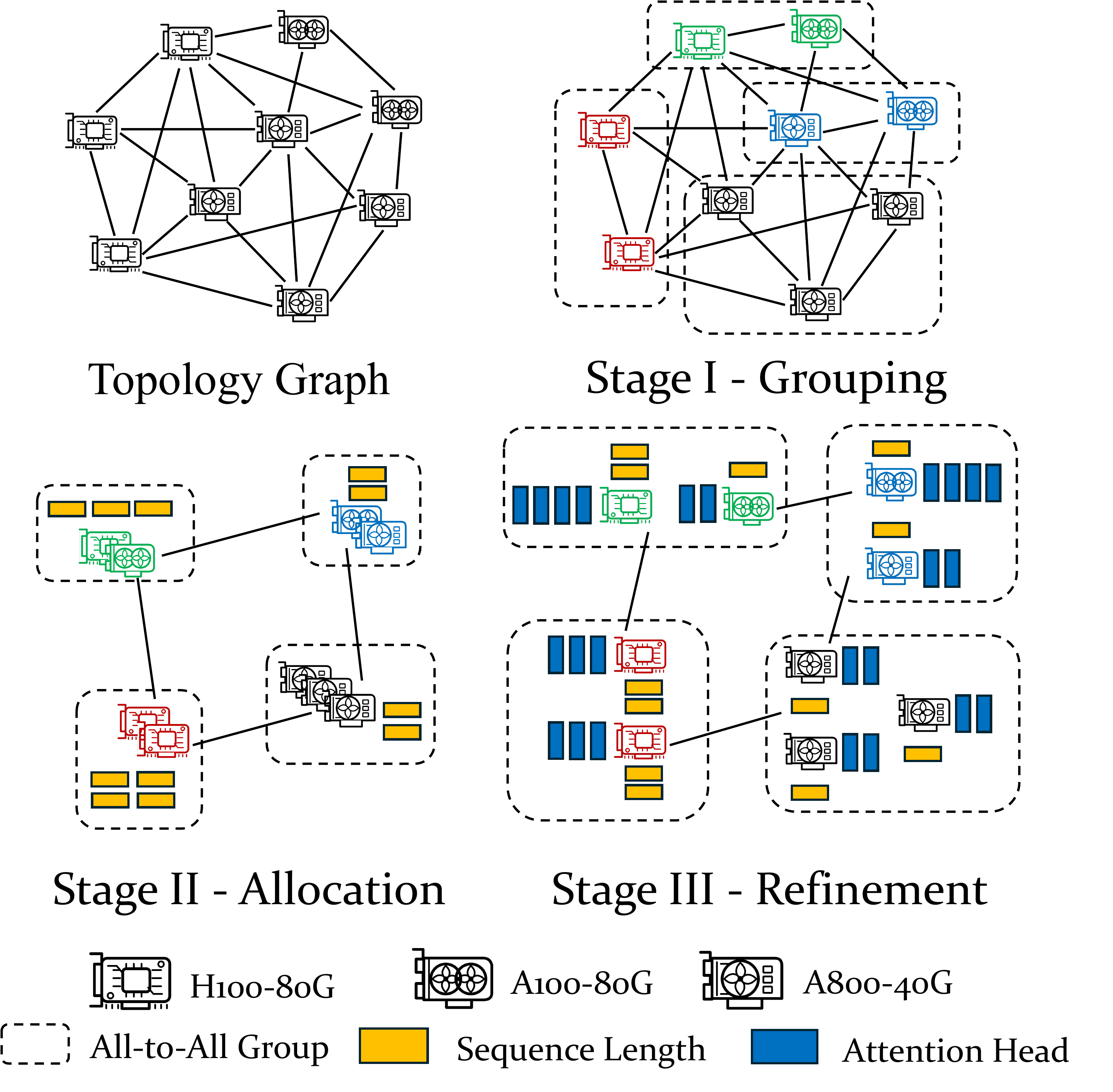}
\caption{Hierarchical scheduling pipeline.}
\label{fig:scheduler_pipeline}
\end{wrapfigure}

\textbf{Overview.}
The schedule space $\mathfrak{T}$ is too large for exhaustive search, as it jointly spans group formation, sequence allocation, and rank-level sequence--head placement. \name therefore uses an efficient three-stage hierarchical scheduler (Figure~\ref{fig:scheduler_pipeline}) that decomposes this coupled search into finer decisions: it first constructs topology-aware A2A groups, then allocates the global sequence length across groups according to their aggregate capabilities, and finally refines rank-level sequence and head placement within each group. The full pseudocode is given in Appendix~\ref{app:scheduler_algorithm}, and we use $M_1=64$, $M_2=16$, and $I_{\max}=100$ throughout the experiments.

\textbf{Stage I: A2A grouping and abstraction.}
Stage~I enumerates candidate A2A partitions $\mathcal{P}$ by greedy agglomeration over the topology graph $\mathcal{G}$, merging high-bandwidth edges first and terminating each merge sequence once the bandwidth gap between successive merges exceeds a threshold. The scheduler keeps the top-$M_1$ partitions ranked by intra-group bandwidth. Each retained group $G_k$ is then summarized as a super-node by its aggregate compute $\sum_{d\in G_k}c_d$, aggregate memory bandwidth $\sum_{d\in G_k}b_d$, conservative memory proxy $\min_{d\in G_k}m_d$, and average communication cost on intra-group and inter-group edges. This abstraction gives the next stage a compact representation of what each A2A group can compute, store, and communicate. Stage~I therefore reduces the device-level partition search to at most $M_1$ super-node summaries, each one a topology-aware seed that the remaining stages refine instead of re-deriving.

\textbf{Stage II: Global sequence partitioning.}
For each retained partition, Stage~II enumerates candidate splits of the global sequence $L_{\mathrm{tot}}$ between groups $\{G_k\}$, subject to $\sum_k L_k=L_{\mathrm{tot}}$. The seed splits are compute-proportional, square-root-compute, and a memory-capped variant that clips a group's $L_k$ whenever it would exceed the activation budget admitted by the memory proxy of that group; small local perturbations around each seed enrich the candidate set. Each candidate is scored by the performance model of \S\ref{subsec:cost_model}, and the top-$M_2$ are forwarded to Stage~III. The output of this stage is therefore a shortlist of group-level plans, each pairing one Stage~I partition with feasible sequence allocation, so Stage~III only refines combinations that already look promising at the coarse granularity.

\textbf{Stage III: Rank-level refinement.}
For each retained group-level plan, Stage~III initializes the per-rank sequence shards $s_d^{\mathrm{pre}}$ and head counts $n_d$, subject to $\sum_{d\in G_k}s_d^{\mathrm{pre}}=L_k$ within each group and $\sum_{d\in\mathcal{D}}n_d=N_{\mathrm{heads}}$ globally. Coordinate descent then proposes moves that transfer a small budget of sequence tokens or attention heads between a pair of devices in the same or adjacent groups; a move is accepted only if it strictly reduces the predicted block latency and the feasibility filter passes for every device. The descent terminates at a local optimum or after a bounded number of iterations, handling integer head counts and the local bottlenecks hidden by the super-node abstraction. Each refined plan is now a fully instantiated schedule whose A2A partition, group-level sequence split, and per-rank sequence and head placement are all decided.

\textbf{Putting it together.}
The hierarchy therefore navigates $\mathfrak{T}$ in a coarse-to-fine manner: it first forms topology-aware groups, then allocates sequence length across groups, and finally refines rank-level sequence and head placement. The early stages prune the candidate space, so the more expensive coordinate descent in Stage~III is applied only to promising plans. Across all expanded plans, \name returns the lowest-cost feasible schedule as $\mathcal{T}^{\star}$.

\section{Experiments}
\label{sec:experiments}
We evaluate \name from three perspectives. First, we measure end-to-end training throughput on physical heterogeneous testbeds (\S\ref{subsec:measured_results}). Second, we use simulation to study larger clusters beyond the available testbeds (\S\ref{subsec:simulated_results}). Third, we compare \name on heterogeneous clusters with the strongest baselines on homogeneous clusters that have comparable aggregate compute (\S\ref{subsec:flop_match}). The measured results are shown in Figures~\ref{fig:thr_grid} and~\ref{fig:thr_13b_long}, and the large-scale simulation results are given in Figure~\ref{fig:thr_sim}.

\subsection{Experimental Setup}
\label{subsec:exp_setup}

\textbf{Models and workloads.}
We evaluate \name on GPT-style decoder-only language models~\citep{shoeybi2019megatron}. The physical testbed experiments cover $3$B, $7$B, and $13$B parameters with context lengths from $8$K to $256$K, and the simulation experiments cover $13$B and $70$B parameters with context lengths up to $1024$K. All runs use a global batch size of $8$. Training inputs are tokenized OpenWebText2 documents~\citep{gao2020pile} packed into fixed-length samples through Megatron-LM~\citep{shoeybi2019megatron,narayanan2021efficient}.

\textbf{Baselines and metric.}
We compare against three standard CP and HP baselines: USP~\citep{fang2024usp}, Ulysses~\citep{jacobs2024system}, and Ring Attention~\citep{liuringattention}. We report throughput in tokens per second (TPS). Among the baselines, USP composes CP and HP into a $2$D mesh whose $(\mathrm{CP},\mathrm{HP})$ factorization of the device count is a tunable hyperparameter. We therefore enumerate all legal $(\mathrm{CP},\mathrm{HP})$ factorizations for USP and report the best throughput. All speedups are computed against the strongest baseline at the same configuration. For each measurement, we discard the first $5$ warmup iterations to absorb CUDA, allocator, and NCCL initialization, and report the mean throughput over the next $10$ stable iterations.

\subsection{End-to-End Testbed Performance}
\label{subsec:measured_results}

We use three mixed H100 and A100 testbeds: \underline{\textit{Setting~1}} pairs $4$ H100s with $4$ A100s, \underline{\textit{Setting~2}} pairs $4$ H100s with $8$ A100s, and \underline{\textit{Setting~3}} pairs $8$ H100s with $8$ A100s. Each GPU generation occupies its own node, with intra-node NVLink at $450$\,GB/s on H100 and $300$\,GB/s on A100, and a cross-node RDMA fabric at $25$\,GB/s.

\textbf{Results and discussions.}
Across all measured configurations, \name achieves a $1.11\times$ mean speedup and a $1.19\times$ peak speedup over the strongest baseline. Since the candidate set of \name always includes the three baseline layouts, the improvement comes from selecting asymmetric schedules rather than excluding baseline-equivalent plans.

\begin{figure}[t!]
\centering
\includegraphics[width=\linewidth]{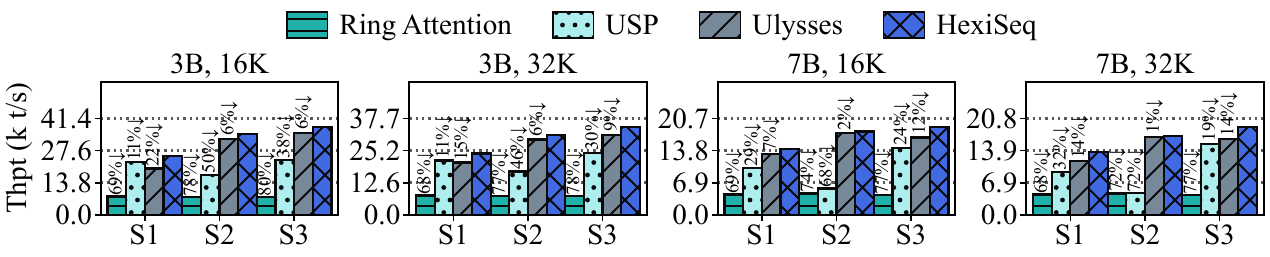}
\caption{End-to-end training throughput of \name compared with three baselines on the $3$B and $7$B models across mixed GPU testbeds.}
\label{fig:thr_grid}
\end{figure}

{\itshape First, {\upshape \name} consistently outperforms the strongest baseline on various model sizes and context lengths.}
As shown in Figure~\ref{fig:thr_grid}, \name achieves the highest throughput at both $16$K and $32$K contexts on the $3$B and $7$B models across all testbed settings. The mean speedup over the strongest baseline rises from $1.08\times$ at $16$K to $1.10\times$ at $32$K; on the $7$B model in Setting~1 the gain grows from $1.08\times$ at $16$K to $1.17\times$ at $32$K, and the same pattern holds on the $7$B model in Setting~3. This trend is consistent with longer attention execution providing additional headroom for asymmetric placement.

\begin{wrapfigure}[13]{r}{0.65\linewidth}
\centering
\vspace{-10pt}
\includegraphics[width=\linewidth]{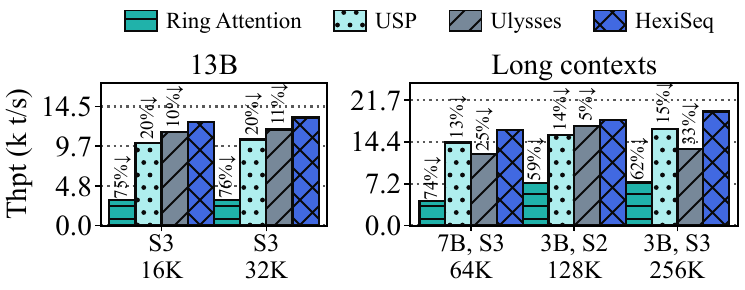}
\caption{Throughput of \name on the $13$B model (left) and three long-context configurations (right).}
\label{fig:thr_13b_long}
\end{wrapfigure}

\textit{Second, the performance advantage remains when the model size increases.}
Figure~\ref{fig:thr_13b_long} reports the $13$B model on Setting~3 with $8$ H100s and $8$ A100s. \name reaches $1.11\times$ the strongest baseline at $16$K and $1.13\times$ at $32$K. The improvement persists despite the higher activation pressure of $13$B, indicating that the asymmetric schedule continues to extract value from the H100 partition once memory headroom tightens.

{\itshape Finally, {\upshape \name} remains effective at the longest contexts that fit on the testbeds.}
Figure~\ref{fig:thr_13b_long} also reports three long-context configurations: $7$B on Setting~3 at $64$K, $3$B on Setting~2 at $128$K, and $3$B on Setting~3 at $256$K. The speedup over the strongest baseline is $1.15\times$ at $64$K, $1.06\times$ at $128$K, and $1.18\times$ at $256$K, with the strongest baseline switching between USP-style and Ulysses-style baselines.

\subsection{Larger-Scale Simulation Performance}
\label{subsec:simulated_results}

\begin{figure}[t!]
\centering
\includegraphics[width=\linewidth]{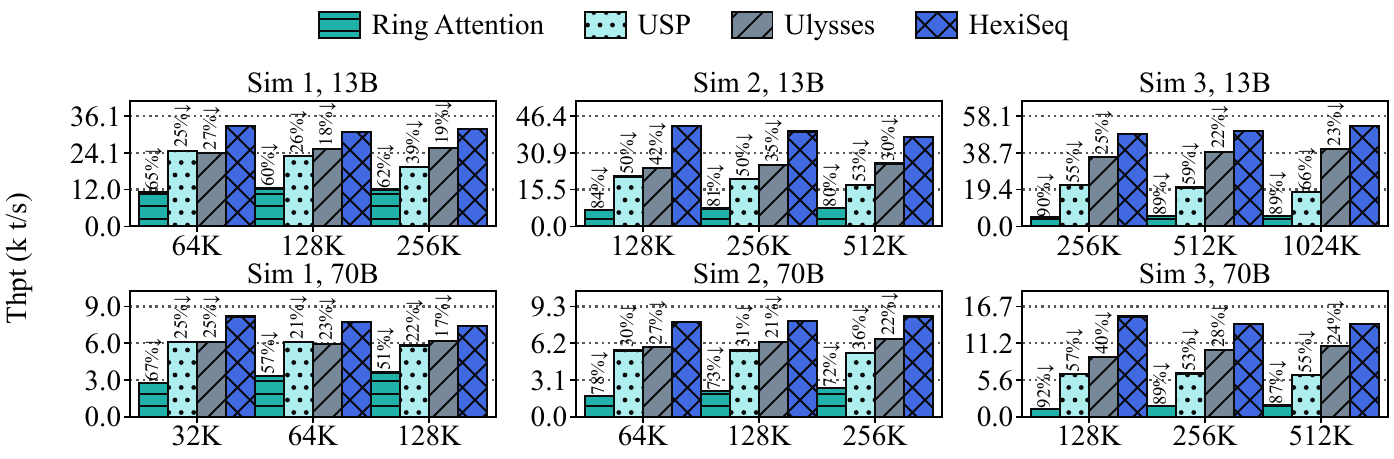}
\caption{Simulated training throughput of \name compared with three baselines on three larger heterogeneous clusters with $13$B (top row) and $70$B (bottom row) models.}
\label{fig:thr_sim}
\end{figure}

To probe scales beyond the physical testbeds, we evaluate \name in simulation using the same scheduler and the cost model. The cost model is calibrated against measured testbed runs: predictions track measured throughput closely and preserve the relative ordering of competing schedules across validation cases (Appendix~\ref{app:cost_validation}), so the simulated throughputs reported below are reliable for comparison at scales we cannot run physically. Intra-node bandwidth varies by accelerator generation: $450$\,GB/s NVLink on H100, $300$\,GB/s on A100, $200$\,GB/s on A800, and $32$\,GB/s PCIe on L40S. All nodes communicate cross-node over a $200$\,GB/s RDMA fabric. We study three heterogeneous clusters with progressively larger scale and GPU diversity: \underline{\textit{Simulation~1}} is a $32$-GPU three-generation cluster with $8$ H100s, $8$ A100s, and $16$ A800s, \underline{\textit{Simulation~2}} is a $64$-GPU four-generation cluster with $16$ each of H100, A100, A800, and L40S, and \underline{\textit{Simulation~3}} is a $128$-GPU four-generation cluster with $16$ H100s, $16$ A100s, $48$ A800s, and $48$ L40S. For each cluster, we report training throughput on $13$B and $70$B language models. Figure~\ref{fig:thr_sim} reports the six model-cluster combinations.

\textbf{Results and discussions.}
{\itshape First, {\upshape \name} achieves the highest throughput on every simulated $13$B configuration.}
As shown in the top row of Figure~\ref{fig:thr_sim}, the largest gain appears on Simulation~2, where the cluster contains four accelerator generations which provide more flexibility for asymmetric placement. \name continues to lead on Simulation~3, the largest simulated cluster in our study.

\begin{wrapfigure}[10]{r}{0.30\linewidth}
\centering
\vspace{-6pt}
\includegraphics[width=\linewidth]{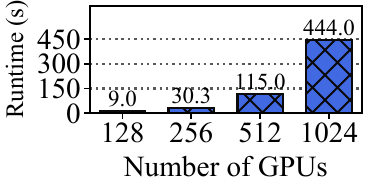}
\caption{Hierarchical scheduler runtime.}
\label{fig:scheduler_eff}
\end{wrapfigure}

\textit{Second, the advantage persists on the $70$B model despite higher memory pressure.}
The bottom row of Figure~\ref{fig:thr_sim} shows that \name remains the strongest method in every panel. Larger models increase memory pressure and per-step communication, but the relative gap does not narrow with context length in these simulations. This result suggests that asymmetric placement remains useful beyond the smaller testbed-scale models.

{\itshape Finally, the scheduler overhead remains manageable at larger scales.} 

 Figure~\ref{fig:scheduler_eff} reports the runtime of the hierarchical scheduler as the number of GPUs increases from $128$ to $1024$. The scheduler finishes in $9.0$ seconds on $128$ GPUs and $444.0$ seconds on $1024$ GPUs, remaining under eight minutes at the largest scale. Thus, the pruning and hierarchical refinement used by \name keep the scheduling overhead manageable compared with the training runs it configures.

\subsection{FLOP-Equivalent Homogeneous Comparison}
\label{subsec:flop_match}

The previous results compare \name with heterogeneous baselines. We further ask whether a heterogeneous cluster can approach the throughput of a homogeneous cluster with similar FLOPs. We study this question in simulation using the cost model from Appendix~\ref{app:cost_validation}. Three baselines run on the homogeneous cluster, while \name runs on the heterogeneous cluster. 

We report two simulated pairs with similar aggregate FLOPs. \underline{\textit{Simulation~4}} pairs $48$ H100s and $8$ A100s on the heterogeneous side with $152$ A100s on the homogeneous side, using a $13$B model at $256$K context. \underline{\textit{Simulation~5}} pairs $52$ H100s, $4$ A100s, and $8$ A800s on the heterogeneous side with $168$ A100s on the homogeneous side, using a $13$B model at $128$K context.

\begin{wrapfigure}[11]{r}{0.6\linewidth}
\centering
\vspace{-8pt}
\includegraphics[width=\linewidth]{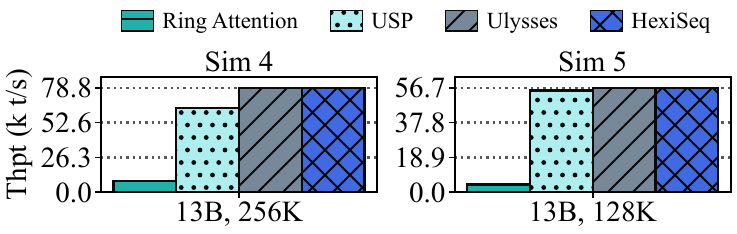}
\caption{FLOP-equivalent comparison of \name and three baselines on heterogeneous clusters.}
\label{fig:thr_flop_match}
\end{wrapfigure}

\textbf{Results and discussions.}
On both FLOP-comparable pairs, the strongest homogeneous baseline is Ulysses. As shown in Figure~\ref{fig:thr_flop_match}, \name stays within $\sim 0.5\%$ of Ulysses throughput on average across Simulation~4 and Simulation~5. With similar aggregate FLOPs, a heterogeneous cluster scheduled by \name therefore reaches throughput close to the strongest homogeneous baseline, indicating that asymmetric placement absorbs most of the heterogeneity penalty under our cost model rather than leaving it as a residual gap.

\section{Conclusion}
\label{sec:conclusion}
\name is a heterogeneous long-context training system that extends context and head parallelism with asymmetric placement and cost-model-guided scheduling algorithm. Empirical studies show that, on mixed H100 and A100 testbeds, \name improves throughput by $1.11\times$ on average and up to $1.19\times$ over the strongest heterogeneous baseline. Simulations on $32$ to $128$ GPUs further show an average gain of $1.36\times$, and FLOP-equivalent comparisons show that \name can approach the throughput of the strongest homogeneous baseline while running on heterogeneous clusters.

%%%%%%%%%%%%%%%%%%%%%%%%%%%%%%%%%%%%%%%%%%%%%%%%%%%%%%%%%%%%

\bibliographystyle{plainnat}
\bibliography{reference}

%%%%%%%%%%%%%%%%%%%%%%%%%%%%%%%%%%%%%%%%%%%%%%%%%%%%%%%%%%%%
\clearpage
\appendix

\section{Detailed Cost Model Derivation}
\label{app:cost_derivation}
This appendix details the memory, computation, and communication terms used by the \name{} performance model of \S\ref{subsec:cost_model}.

\begin{table}[H]
\centering
\caption{Additional notation used in the cost-model derivation. Symbols already defined in Table~\ref{tab:notation} are reused without redefinition.}
\label{tab:cost_model_notation}
\small
\setlength{\tabcolsep}{3pt}
\renewcommand{\arraystretch}{0.96}
\begin{tabularx}{\linewidth}{@{}l | X | l | X@{}}
\hline
\textbf{Symbol} & \textbf{Description} & \textbf{Symbol} & \textbf{Description} \\
\hline
$B$ & Batch or microbatch size & $H$ & Model hidden dimension \\
\hline
$d_h$ & Hidden dimension per attention head & $P_{\mathrm{dtype}}$ & Bytes per tensor element \\
\hline
$G(d)$ & A2A group containing device $d$ & $G_{\mathrm{src}}(t)$ & Source group at Ring step $t$ \\
\hline
$L_{G(d)}$ & Sequence length of $d$'s group ($L_k$ for $d\in G_k$) & $\gamma_{\mathrm{act}}$ & Activation-retention coefficient \\
\hline
$\mathrm{peer}(d,t)=u$ & Source peer of device $d$ at Ring step $t$ & $\mathrm{Msg}(d,t)$ & Ring message size at device $d$ and step $t$ \\
\hline
$\alpha_{i,j}$ & Startup latency between $i$ and $j$ & $\beta_{i,j}$ & Per-byte communication cost ($\beta_{i,j}=1/B_{i,j}$) \\
\hline
$F_{\mathrm{nonattn}},F_{\mathrm{attn}}$ & Non-attention and attention FLOPs & $F_{\mathrm{comp\mbox{-}step}}$ & Per-step attention FLOPs \\
\hline
$V_{\mathrm{nonattn}}$ & Non-attention data movement & $V_{d\to j}$ & A2A volume from $d$ to $j$ \\
\hline
$T_{\mathrm{blk}}$ & Block latency & $T_{\mathrm{iter}}$ & Iteration latency \\
\hline
$T_{\mathrm{nonattn}}$ & Non-attention latency & $T_{\mathrm{A2A}}$ & A2A redistribution latency \\
\hline
$T_{\mathrm{step}}$ & Ring Attention step latency & $\mathrm{Mem}_{\mathrm{act}}(d;\mathcal{T})$ & Peak activation memory estimate \\
\hline
\end{tabularx}
\end{table}

\subsection{Counting Conventions}
\label{app:counting}

Equation~\eqref{eq:total_time} omits explicit execution multiplicities; forward, backward, and recomputation effects are absorbed into the per-component FLOP and communication-volume formulas. Non-attention modules (QKV projections, MLP, normalization) include forward and backward execution; attention computation includes forward, backward, and activation recomputation; communication costs include forward and backward execution, while recomputation does not incur additional communication.

\subsection{Memory Cost Model}

To ensure feasibility, the peak memory usage on each device $d$ must not exceed its capacity $m_d$. We decompose $\mathrm{Mem}(d;\mathcal{T})$ into a schedule-independent residency $M^{\mathrm{static}}_d$ that absorbs weights, gradients, and optimizer states under the chosen DP/PP/TP outer layout, and a schedule-dependent activation peak $\mathrm{Mem}_{\mathrm{act}}(d;\mathcal{T})$ that varies with $s_d^{\mathrm{pre}}$, $L_{G(d)}$, and $n_d$. The feasibility filter then applies $M^{\mathrm{static}}_d + \mathrm{Mem}_{\mathrm{act}}(d;\mathcal{T}) \le m_d$, so schedule scoring reduces to estimating $\mathrm{Mem}_{\mathrm{act}}(d;\mathcal{T})$. We estimate the activation peak by summing the dominant saved tensors required for backward and recomputation under our implementation. Each device stores (i) the non-attention activation stream before the sequence-to-head redistribution, of shape $[B, s_d^{\mathrm{pre}}, H]$, and (ii) the attention KV staging buffers used by Ring Attention, of shape $[B, L_{G(d)}, n_d\cdot d_h]$ for both $K$ and $V$. Therefore,
\begin{equation}
\label{eq:mem_act_shapes}
\mathrm{Mem}_{\mathrm{act}}(d;\mathcal{T})
\approx
B \cdot P_{\mathrm{dtype}} \cdot
\Big(
\gamma_{\mathrm{act}} \cdot s_d^{\mathrm{pre}} \cdot H
+
2 \cdot L_{G(d)} \cdot (n_d\cdot d_h)
\Big),
\end{equation}
where $\gamma_{\mathrm{act}}$ is the number of backbone activation tensors of shape $[B, s_d^{\mathrm{pre}}, H]$ that must be retained at peak (e.g., residual stream and normalization inputs and outputs) under the chosen checkpointing and fusion strategy. In our implementation we set $\gamma_{\mathrm{act}}=2$.

\subsection{Computation Cost Model}

We model the cost of a full training iteration, including forward, backward, and recomputation phases.

\textbf{Non-attention computation.}
This component includes QKV and output projections, MLP layers, and normalization. The FLOPs and memory traffic on device $d$ scale with its pre-A2A sequence shard $s_d^{\mathrm{pre}}$. We approximate the activation traffic of non-attention modules by counting the read and write volume of their input and output activation tensors. Under a standard decoder-only Transformer block with MLP expansion ratio $4$, this yields an effective coefficient of about $20$ per forward pass, or $40$ when accounting for both forward and backward execution:
\begin{equation}
    V_{\mathrm{nonattn}}(d)
    \approx
    40 \cdot B \cdot s_d^{\mathrm{pre}} \cdot H \cdot P_{\mathrm{dtype}}.
\end{equation}
We approximate the FLOPs of these modules as
\begin{equation}
    F_{\mathrm{nonattn}}(d) \approx 72 \cdot B \cdot s_d^{\mathrm{pre}} \cdot H^2,
\end{equation}
which aggregates the dominant dense GEMMs in one Transformer block under the standard MLP expansion ratio. We then use a roofline model to estimate latency:
\begin{equation}
\label{eq:nonattn_time}
T_{\mathrm{nonattn}}(d) =
\max \left(
\frac{F_{\mathrm{nonattn}}(d)}{c_d},
\frac{V_{\mathrm{nonattn}}(d)}{b_d}
\right).
\end{equation}

\textbf{Attention computation.}
Device $d$ computes attention for its assigned heads $n_d$ over its local query block $L_{G(d)}$ against the global context. The total floating-point operations, aggregated across all Ring Attention steps, are
\begin{equation}
    F_{\mathrm{attn}}(d) \approx
    16 \cdot B \cdot L_{G(d)} \cdot (n_d \cdot d_h) \cdot L_{\mathrm{tot}}.
\end{equation}
The latency of attention is not modeled as a standalone term; instead, it is decomposed into per-step execution within the Ring Attention pipeline, where computation and communication overlap.

\subsection{Communication Cost Model}

\textbf{Heterogeneous A2A.}
The A2A operation redistributes tensors between the sequence and head dimensions. In contrast to homogeneous settings, the model tracks the exact point-to-point data volume sent from device $d$ to device $j$ under the relaxed partitioning. The volume $V_{d \to j}$ corresponds to the slice of $d$'s sequence shard $s_d^{\mathrm{pre}}$ destined for device $j$'s assigned heads:
\begin{equation}
    V_{d \to j} = 3 \cdot B \cdot s_d^{\mathrm{pre}} \cdot (n_j \cdot d_h) \cdot P_{\mathrm{dtype}}.
\end{equation}
The factor $3$ accounts for $Q$, $K$, and $V$ in the forward redistribution. Each Transformer block invokes the A2A primitive four times (two in the forward pass and two in the backward pass); applying the QKV volume to all four yields a conservative high-volume estimate, since the reverse redistributions may carry fewer tensors depending on the implementation. The latency of group $G_k$ is bottlenecked by the slowest point-to-point link in the group:
\begin{equation}
\label{eq:a2a_latency}
T_{\mathrm{A2A}}(G_k) =
4 \cdot
\max_{d \in G_k}
\max_{j \in G_k, j \neq d}
\left(
\alpha_{d,j}
+
\beta_{d,j} \cdot V_{d \to j}
\right).
\end{equation}

\textbf{Heterogeneous Ring Attention.}
The Ring Attention pipeline executes in $K$ steps ($t=0,\ldots,K-1$). Let $u = \mathrm{peer}(d, t)$ denote the source device from which device $d$ receives data at step $t$. We distinguish between the local attention step (where source and destination groups coincide) and the remote communication steps. For the local step (typically $t=0$), no communication occurs, so $\mathrm{Msg}(d,0) = 0$. For remote steps ($t > 0$), the message size depends on the sequence length held by the source group $G_{\mathrm{src}}(t)$:
\begin{equation}
    \mathrm{Msg}(d,t) =
    4 \cdot B \cdot L_{G_{\mathrm{src}}(t)} \cdot (n_d \cdot d_h) \cdot P_{\mathrm{dtype}}.
\end{equation}
The computation load at step $t$ corresponds to the blockwise attention between the local query block $L_{G(d)}$ and the source KV block $L_{G_{\mathrm{src}}(t)}$:
\begin{equation}
    F_{\mathrm{comp\mbox{-}step}}(d,t)
    \approx
    16 \cdot
    B \cdot
    L_{G(d)} \cdot
    L_{G_{\mathrm{src}}(t)} \cdot
    (n_d \cdot d_h).
\end{equation}
Assuming overlap between communication and computation, the latency of step $t$ is
\begin{equation}
\label{eq:ring_step}
T_{\mathrm{step}}(t) =
\max_{d \in \mathcal{D}}
\begin{cases}
    \dfrac{F_{\mathrm{comp\mbox{-}step}}(d,t)}{c_d} & \text{if } t=0 \text{ (local)}, \\[6pt]
    \max \left(
        \dfrac{F_{\mathrm{comp\mbox{-}step}}(d,t)}{c_d}, \;
        \alpha_{u,d} + \beta_{u,d} \cdot \mathrm{Msg}(d,t)
    \right) & \text{if } t > 0 \text{ (remote)}.
\end{cases}
\end{equation}

\section{Hierarchical Scheduler Algorithm}
\label{app:scheduler_algorithm}
Algorithm~\ref{alg:scheduler} gives the pseudocode of the hierarchical scheduler described in \S\ref{subsec:scheduler}. Stage~I (lines~1--2) constructs candidate A2A partitions by topology-aware graph agglomeration and summarizes each retained group as a super-node. Stage~II (line~3) generates candidate splits of the global sequence across groups. Stage~III (lines~4--19) iterates over each retained group-level plan, initializes the rank-level assignment, and improves it with bounded-iteration coordinate descent. The candidate generators \textsc{PartitionTopology}, \textsc{AbstractSuperNodes}, \textsc{GenerateCandidates}, \textsc{InitializeAssignment}, \textsc{CandidatePairs}, and \textsc{ProposeMove} are described in \S\ref{subsec:scheduler}; \textsc{CostModel} and \textsc{FeasibilityCheck} invoke the analytical performance model of \S\ref{subsec:cost_model}.

\begin{algorithm}[t]
\renewcommand{\algorithmicrequire}{\textbf{Input:}}
\renewcommand{\algorithmicensure}{\textbf{Output:}}
\caption{Hierarchical scheduler for heterogeneous CP and HP.}
\label{alg:scheduler}
\small
\begin{algorithmic}[1]
\Require Devices $\mathcal{D}$ with profiles $\{c_d, b_d, m_d\}_{d\in\mathcal{D}}$, topology graph $\mathcal{G}$, sequence length $L_{\mathrm{tot}}$.
\Ensure Schedule $\mathcal{T}^{\star}$.
\State $\mathcal{P}_{\mathrm{A2A}} \gets \textsc{PartitionTopology}(\mathcal{G})$ \Comment{Stage I: candidate topologies}
\State $\mathcal{S} \gets \textsc{AbstractSuperNodes}(\mathcal{P}_{\mathrm{A2A}})$ \Comment{Stage I: super-node abstraction}
\State $\mathcal{W} \gets \textsc{GenerateCandidates}(\mathcal{S}, L_{\mathrm{tot}})$ \Comment{Stage II: group sequence planning}
\State $\mathcal{T}^{\star} \gets \varnothing$;\quad $C^{\star} \gets \infty$
\ForAll{$\mathbf{w} \in \mathcal{W}$}
    \State $\mathcal{T} \gets \textsc{InitializeAssignment}(\mathbf{w}, \mathcal{S})$ \Comment{Stage III: rank-level refinement}
    \State $C \gets \textsc{CostModel}(\mathcal{T})$
    \State $\textit{improved} \gets \textbf{true}$;\quad $i \gets 0$
    \While{$\textit{improved}$ \textbf{and} $i < I_{\max}$}
        \State $\textit{improved} \gets \textbf{false}$;\quad $i \gets i + 1$
        \ForAll{$(d_a, d_b) \in \textsc{CandidatePairs}(\mathcal{T})$}
            \State $\mathcal{T}' \gets \textsc{ProposeMove}(\mathcal{T}, d_a, d_b)$
            \If{\textsc{FeasibilityCheck}($\mathcal{T}'$)}
                \State $C' \gets \textsc{CostModel}(\mathcal{T}')$
                \If{$C' < C$}
                    \State $\mathcal{T} \gets \mathcal{T}'$;\quad $C \gets C'$;\quad $\textit{improved} \gets \textbf{true}$
                    \State \textbf{break}
                \EndIf
            \EndIf
        \EndFor
    \EndWhile
    \If{$C < C^{\star}$}
        \State $\mathcal{T}^{\star} \gets \mathcal{T}$;\quad $C^{\star} \gets C$
    \EndIf
\EndFor
\State \Return $\mathcal{T}^{\star}$
\end{algorithmic}
\end{algorithm}

\section{Cost-Model Validation on the Testbeds}
\label{app:cost_validation}
\name's scheduler ranks candidate schedules using the analytical performance model in \S\ref{subsec:cost_model}. Therefore, the relevant validation criterion is whether the model predicts both absolute throughput within a reasonable range and, more importantly, the relative ordering among competing schedules. Table~\ref{tab:prediction_gap_summary} reports the per-method prediction gap, defined as $(\mathrm{simulated}-\mathrm{measured})/\mathrm{measured}$, on a representative subset of validation cases covering all three baselines and the schedules selected by \name on the testbeds in \S\ref{subsec:exp_setup}. Most points fall within $\pm 10\%$, and the model's relative ordering of schedules matches the measured ordering in every case shown. Although larger absolute outliers appear in the full evaluation set, they do not change the ranking between \name and the strongest baseline at the same configuration. This ranking fidelity justifies using the same simulator to compare schedules at scales beyond the physical testbeds in \S\ref{subsec:simulated_results}.

\begin{table}
\centering
\small
\begin{tabular}{llrrrr}
\toprule
Method & Model/Setting & Context & Measured (kTPS) & Simulated (kTPS) & Gap \\
\midrule
USP & 3B, Setting~2 & 64K & 16.6 & 16.8 & $+1.6\%$ \\
USP & 3B, Setting~3 & 64K & 27.4 & 29.9 & $+9.3\%$ \\
Ulysses & 3B, Setting~3 & 128K & 19.8 & 19.0 & $-4.1\%$ \\
Ulysses & 13B, Setting~3 & 16K & 11.4 & 11.2 & $-1.5\%$ \\
Ring & 3B, Setting~2 & 32K & 7.3 & 7.6 & $+4.1\%$ \\
Ring & 7B, Setting~3 & 16K & 4.3 & 4.3 & $+1.5\%$ \\
\name & 3B, Setting~3 & 64K & 30.2 & 30.0 & $-0.6\%$ \\
\name & 7B, Setting~3 & 64K & 16.5 & 15.9 & $-3.8\%$ \\
\bottomrule
\end{tabular}
\caption{Cost-model prediction gap on a representative validation subset measured on the mixed H100/A100 testbeds of \S\ref{subsec:exp_setup}. Throughputs are reported in $10^3$ tokens/s; the gap is $(\text{simulated}-\text{measured})/\text{measured}$. The full validation set covers 76 cases; the rows below are illustrative across methods, models, settings, and context lengths.}
\label{tab:prediction_gap_summary}
\end{table}

\section{Limitations}
\label{app:limitations}
\name focuses on heterogeneous CP and HP scheduling for long-context attention. Our implementation and evaluation cover the attention-parallel part of the training stack, while leaving broader cluster-level placement decisions to future work.

\textbf{Evaluation scale.}
Our direct measurements use mixed H100 and A100 testbeds of up to $16$ GPUs, covering $3$B, $7$B, and $13$B models with context lengths from $8$K to $256$K. To study larger regimes, we use the validated performance model from Appendix~\ref{app:cost_validation} to simulate $32$ to $128$ GPU clusters, $13$B and $70$B models, and contexts up to $1024$K. This combination lets us validate \name end to end on real heterogeneous hardware and then isolate scaling trends beyond the available testbeds. Future large-cluster measurements would further quantify deployment-specific effects such as background traffic and NCCL protocol choices.

\textbf{Scheduling scope.}
\name optimizes the CP and HP attention schedule while treating data, tensor, and pipeline parallelism as surrounding decisions. This design keeps the schedule space tractable and allows \name to be composed with existing training stacks. A natural extension is to co-schedule these outer parallelism dimensions together with sequence and head placement, enabling a single optimizer to reason about model shards, microbatches, and attention partitions jointly.

%%%%%%%%%%%%%%%%%%%%%%%%%%%%%%%%%%%%%%%%%%%%%%%%%%%%%%%%%%%%

\clearpage

\end{document}